# Diminish Electrostatic in Piezoresponse Force Microscopy through longer ultra-stiff tips

A. Gómez[a]*, M. Coll[a], T. Puig[a] and X. Obradors[a]

[a]Institut de Ciència de Materials de Barcelona (ICMAB-CSIC), Campus UAB, Bellaterra, 08193, Catalonia, Spain

## ABSTRACT

Piezoresponse Force Microscopy is a powerful but delicate nanoscale technique that measures the mechanical response resulting from the application of a highly localized electric field. Though mechanical response is normally due to piezoelectricity, other physical phenomena, especially electrostatic interaction, can contribute to the signal read. We address this problematic through the use of longer ultra-stiff probes providing state of the art sensitivity, with the lowest electrostatic interaction and avoiding working in high frequency regime. In order to find this solution we develop a theoretical description addressing the effects of electrostatic contributions in the total cantilever vibration and its quantification for different setups. The theory is subsequently tested in a Periodically Poled Lithium Niobate (PPLN) crystal, a sample with well-defined 0º and 180º domains, using different commercial available conductive tips. We employ the theoretical description to compare the electrostatic contribution effects into the total phase recorded. Through experimental data our description is corroborated for each of the tested commercially available probes. We propose that a larger probe length can be a solution to avoid electrostatic forces, so the cantilever-sample electrostatic interaction is reduced. For hard oxide samples we propose an ultra-stiff cantilever which avoids the use of high frequency voltage but still diminishing electrostatic forces. Our proposed commercially available solution have great implications into avoiding artifacts while studying soft biological samples, multiferroic oxides, and thin film ferroelectric materials and it opens a new window into tip engineering.

## 1. Introduction

Piezoresponse Force Microscopy (PFM) is a powerful tool based on Atomic Force Microscopy (AFM) to probe piezo and ferroelectric properties, at the nanoscale, in a wide variety of materials ranging from thin film complex oxides to biological samples[1–8]. In the PFM method, a conductive AFM tip is used in contact mode operation, while an AC Voltage signal is directly applied to the tip. As a result of the converse piezoelectric effect, the material vibrates at the same frequency as the AC voltage used. The vibration is then transmitted to the cantilever and recorded by the AFM photo-detector as a deflection signal, giving information about the piezoelectric properties of the material[9–12].

Even though the method is simple, the interpretation of the resulting signal is not straightforward as it has a great dependence on external forces proportional to the AC voltage used[13,14]. Anomalous behaviours in PFM images have been observed while studying non-ferroelectric materials[15] and solutions to avoid data misleading interpretation have been proposed[16]. Furthermore, the on-going solutions include methods to separate the electrostatic contribution from piezoelectric contribution[17], data mining[18], interferometer sensors as a replace to optical beam deflections systems[19], among others.

However, some challenges still remain, like diminishing the electrostatic forces from pure piezoelectric forces, which is crucial for studying thin film ferroelectric materials. In this work we tackle the problem through a new perspective, as our target is to diminish electrostatic signal, rather than differentiating piezoelectric from electrostatic signals. We study the effect of an external force proportional to the AC voltage, applying a Lock-in Amplifier (LIA) based theoretical model. The conclusions from this novel approach are tested in a well-known ferroelectric sample as a Periodically Poled Lithium Niobate (PPLiNbO$_3$)[20]. The theory is used to improve the PFM signal through the study of different commercial available conductive tips, finding the one that minimizes electrostatic contribution for PFM measurements, while maintaining state of the art sensitivity. We find the lowest electrostatic influenced system through the use of longer ultra-stiff tips, resonating below megahertz range, which are commercially available and usable in any AFM equipment.

## 2.    Theoretical Approach

The electromechanical behaviour of the AFM tip in PFM experiments has been found to be related to different force interactions, being the most important ones piezoresponse and electrostatic. The electrostatic effect have been studied previously by other authors, finding that it can seriously mislead the interpretation of data as, for instance, reading a ferroelectric signal in a non-ferroelectric material[13,21]. As electrostatic is such an important factor, we developed a theoretical description to explain the effect of a small electrostatic signal in the piezoresponse signal. We assume that the AFM system reads two cantilever oscillations of the same frequency, different amplitudes and different phases. The oscillations are represented as electronic signals, being one of them the vibration of the cantilever due to piezoelectric forces, while the other signal is the movement of the cantilever due to electrostatic interactions[22]. Both signals can be mathematically represented as a sum of individual signals:

$$A(t) = A_{Piezo} \sin(\omega t + \emptyset_{Piezo}) + A_{Elec} \sin(\omega t + \emptyset_{Elec}) \quad (1)$$

Where $A_{Piezo}$, $\emptyset_{Piezo}$ are the amplitude and phase of vibration due to Piezoelectric forces; $A_{Elec}$, $\emptyset_{Elec}$ are the Amplitude and Phase of vibration due to electrostatic forces and ω is the driving frequency of the AC Voltage applied to the tip. We introduce both signals into a LIA, to calculate the Phase of the whole signal, taking into account both piezoresponse and electrostatic interactions. The LIA uses the correlation function to obtain the amplitude and phase parameters of a given sinusoidal function[23]. We can use the correlation function with (1)

$$R(\delta, T) = \frac{1}{T} \int_0^T f(t) A(t + \delta) dt \quad (2)$$

Where T is the upper integration limit, f(t) is the reference signal, A(t) is the modulated response and δ is a time parameter. Using the integral properties we find

$$R(\delta, T) = \frac{1}{T} \int_0^T \sin(\omega t) A_{Piezo} \sin(\omega t + \emptyset_{Piezo}) \, dt + \frac{1}{T} \int_0^T \sin(\omega t) A_{Elec} \sin(\omega t + \emptyset_{Elec}) dt \quad (3)$$

The two independent integrals can be solved using standard Lock-in amplifiers theory[24]. Operating the integrals, it is found that the phase seen by the LIA has the following expression:

$$\text{Phase} = \tan^{-1}\left(\frac{U_1}{U_2}\right) \quad (4)$$

Where

$$U_1 = A_{Piezo}\cos(\emptyset_{Piezo}) + A_{Elec}\cos(\emptyset_{Elec}) \quad (5a)$$
$$U_2 = A_{Piezo}\sin(\emptyset_{Piezo}) + A_{Elec}\sin(\emptyset_{Elec}) \quad (5b)$$

the formula (4) is the expression of the phase for the whole system, including both piezoelectric and electrostatic interactions. According to the expression (4), it is shown that $A_{Elec}$ is a parameter that can change the phase of the whole system which misleads the interpretation of acquired data.

The fact that an electrostatic signal can mislead the interpretation have been reported experimentally[25]. However, a full theoretical description has to be developed, in order to understand the effects of this contribution, which is crucial, and it is the main purpose of this novel description[15,17,25–31]. Minimizing the electrostatic contribution is essential to have a true piezoresponse signal. To study the contribution of $A_{Elec}$ into the total phase recorded by the LIA, we apply this theory to describe an ideal sample that only has domains poled up or down. We assume that the electrostatic force is a constant phase force, because the DC voltage is always maintained in 0V[32,33]. Mathematically, the conditions are as follow

$$\emptyset_{PiezoUP} = 0º, \emptyset_{PiezoDOWN} = 180º, \emptyset_{Elec} = 30º$$

where $\emptyset_{PiezoUP}$, $\emptyset_{PiezoDOWN}$ are the phase of the up and down domains of the sample and $\emptyset_{Elec}$ is the phase of the electrostatic forces. A phase of 30º was selected because it should be positive, thus attractive force, and it is a common value when measuring electrostatic forces with AFM, which ranges from 5º to 70º. With the following conditions, we can study the resulting phase difference of domains up and down of the whole system as a function of the ratio between $A_{Elec}/A_{Piezo}$. In **Figure 1**, it is shown the phase of the up and down domains as well as the phase difference between both domains, versus the relation $A_{Elec}/A_{Piezo}$. It is seen that as the $A_{Elec}/A_{Piezo}$ ratio increases, the phase difference between up and down domain decreases. Moreover, the phase change between up and down domains is not symmetric. This asymmetric contribution is related to the difference between the electrostatic signal phase and the Piezoresponse phase for each of the domains. The electrostatic phase, 30º, is closer to the up domain phase, which is 0º, and hence its contribution is lower as compared to the down domain case.

Here we select three representative situations to emphasize the relevance of the $A_{Elec}/A_{Piezo}$ ratio. For high electrostatic contribution, $A_{Elec}/A_{Piezo}$ equals to 2 or greater, the domains cannot be seen and, moreover, the global phase of each domain is proportional to the electrostatic phase signal, see **Figure 2**. As a consequence, a change of electrostatic phase signal changes the resulting total phase seen by the LIA, independently of the piezoresponse signal phase. This behaviour could mislead the interpretation of a non-ferroelectric material as being ferroelectric. For the low electrostatic contribution regime, $A_{\text{Elec}}/A_{\text{Piezo}}$ is 0.001, the domains have a homogeneous 180º phase difference; the majority of the signal comes from the piezoresponse signal. For $A_{\text{Elec}}/A_{\text{Piezo}}$= 0.5, the total phase difference of the domains has its minimum at $\emptyset_{\text{Elec}} = 90º$, where the phase measured is 126º, diminishing the phase difference being read between domains.

## 3. Experimental Results

At this step, we employed the described theory to experimentally illustrate the relevance of $A_{\text{Elec}}/A_{\text{Piezo}}$. We studied PPLN crystal, which presents well-defined 0º and 180º domains, using different commercial available conductive tips (see **Table 1**). For PFM measurements, we have selected stiff cantilevers rather than soft cantilevers, to reduce electrostatic effects[34]. A stiff cantilever has a lower thermal noise, however, in rectangular cantilevers, a stiffer cantilever will drop its sensitivity and increase the contact resonance frequency to the Megahertz range[35]. For Rocky Mountain Nanotechnology (RMN) tips, neither deflection sensitivity nor resonance frequency are related to the k constant of the cantilever. There is also an outstanding advantage of RMN tips, the larger tip length (see S1 in Supplementary Information). Such a large tip length greatly reduces the cantilever-sample capacitive coupling and therefore the electrostatic contribution. Larger tip length probes have been custom manufactured and employed for imaging biological samples, but the use of a larger tip length to avoid electrostatic interactions is a solution which was not implemented, tested and demonstrated previously[36]. Another advantage of using RMN tips is the aspect ratio of the tip, as being sharper, the capacitive coupling between the tip and the sample is also reduced[37]. At this point, the test sample is scanned with three conductive tips, all of them commercially available, which are NanoWorld EFM (NW EFM), RockyMountain Nanotechnology 25PT400 (RMN-25PT400) and RockyMountain Nanotechnology 25PT200H (RMN-25PT200H-B). We specifically selected these probes as the first one is a platinum coated tip, which is the most common in PFM experiments, while the second is a larger tip length probe. Both platinum coated tip and longer tip length probe have a similar spring constant, in order to discard the stiffness of the cantilever as a variable. The third probe, RMN-25PT200H-B consists of an ultra-stiff probe, with spring constant of 250 N/m, maintaining good deflection sensitivity and a contact resonance frequency of 240 kHz.

We scanned a 10x10 microns area, located in the same spot of the sample, to minimize possible sample in-homogeneities. A similar force was applied between tip and sample to exclude the force as a variable[14]. The scanned area was selected between two opposite domains of the sample. The AC voltage frequency was kept constant at 105 kHz. We choose this frequency because it is far from the contact resonance frequency of any of the three tips, so possible resonant artefacts are avoided. We started the image applying 5 VAC, and we increased the AC voltage amplitude in 1V steps, each 2 microns, until a maximum of 9 VAC was achieved, see **Figure 3**. The electrostatic forces tend to saturate at higher voltages, so during this image, we can assume that the electrostatic forces became constant from 5VAC to 9VAC. With this assumption, we can interpret the contrast change in the image as an increase in the piezoelectric signal over the electrostatic signal, each time the applied AC voltage amplitude is increased. It is observed that by increasing the AC voltage amplitude, the phase difference between the up and down domains is also increased (see S2 in SI). The proposed methodology was tested using two different tips, with the exact same part number, in order to test the reproducibility of the results (see S3 in SI). Furthermore, the increased phase is not the same for up and down domain phases, an asymmetry is found (see S4 in SI). This asymmetry was not previously explained; however the proposed theoretical description explains this experimental result, as the electrostatic signal interacts greater with the down domain phase, compared to the up domain.

Through this experimental method, we can now compare which of the setups provides a piezoelectric signal less influenced by electrostatic forces. The ideal case consists of reading a full 180º phase signal between domains, independently of the AC voltage used. The experimental data was fitted in the theoretical values of **Figure 1**, where fitting is shown in **Figure 4** for a $\emptyset_{\text{Elec}}$ of 9º. This particular value was used to improve the experimental data fitting of the four tips used. Within these conditions, we find that the electrostatic contribution of the longer tip is 1.9 times lower than standard platinum tip and 1.5 times lower than a diamond coated tip (NW CDT-FMR)(see S5 of SI). For ultra-stiff tips, we find that the electrostatic contribution is 2.9 times lower compared to standard platinum tip and 2.03 compared to diamond coated tip. Even though the electrostatic contribution cannot be fully separated from piezoresponse signal, its contribution can be greatly diminished by optimizing the setup used.

Finally, we performed another experiment to corroborate our previous findings. We change the phase of the electrostatic signal phase by applying a DC voltage to the sample. We used a diamond doped DD-SICONA probe, with very low k constant cantilever of 0.2N/m, which is extremely sensitive to little forces as Electrostatic ones, as well as cantilever buckling[14]. PFM phase image, **Figure 5a**, of the test sample, were acquired with an AC Voltage of 5V amplitude while applying a DC bias of +5 V from its bottom to the middle part and a DC bias of -5VDC from the middle to the upper part. As denoted by PFM images, (see S6 of SI) the ferroelectric domains are vertically aligned, however the piezoelectric signal is completely overlapped by electrostatic contribution. We can use our theory to explain this result, as we are now in the regime where $A_{\text{Elec}}/A_{\text{Piezo}}$ is greater than 2, where a change of 180º of $\emptyset_{\text{Elec}}$ shifts the phase of the system by 180º. This phase shift is not related to domain polarization due to the electric field under the tip, as we are far away from the coercive field of the crystal[27,38]. The same experiment was repeated with other three tips, a standard platinum coated tip, **Figure 5b**, a longer tip, **Figure 5c** and a ultra-stiff tip, **Figure 5d**. It is found that the less influenced by electrostatic properties is the ultra-stiff tip, while the longer tip length probe provides a less electrostatically influenced signal, compared with its counterpart, the standard platinum coated tip. Through this experiment, we can confirm our proposed solution to diminish the electrostatic contribution by classifying the tips as a function of $A_{\text{Elec}}/A_{\text{Piezo}}$ ratio.

## 4. Conclusions

We have proposed a theoretical description to compare the effect of electrostatic forces in Piezoresponse Force Microscopy (PFM) for different setups. The theoretical description is subsequently validated through a series of proposed experiments in order to experimentally quantify the electrostatic contribution. It is found that conventional tips used for PFM could mislead the PFM phase results interpretation due to electrostatic effects. A method, based in scanning a Periodically Polled Lithium Niobate (PPLiNbO$_3$) is presented in order to quantify and compare the electrostatic contribution of each tip. The method proposed does not need any special equipment or modified setup. After studying commercially available conductive tips for PFM, we compare the electrostatic influence of each tip, finding the tip which minimizes the electrostatic contribution, maximizes deflection sensitivity value and which resonates below the MHz range, providing state of the art PFM measurements. Our proposed solution, using longer ultra-stiff tips can be immediately implemented in any AFM setup without any physical modification avoiding the use of high frequency piezoresponse.

## Experimental setup

The test sample is Periodically Poled Lithium Niobate (PPLN), a sample that has only domains up and down, with a domain size of 10-50 um width and 400 um length. The sample, which is commercially available, has a piezoelectric coefficient d33=7.5 pm/V, coercive field 2x10$^7$ V/m, surface polarization 0.7 C/m$^2$ [38]. The equipment used is an Agilent 5500 SPM with the AC Mode III accessory. The AC frequency is 105 kHz for all PFM measurements, out of resonance, to avoid possible resonant artefacts. All the images were acquired in low ambient humidity conditions, less than 8%.

## Acknowledgements

This research was supported by Consolider NANOSELECT (CSD 2007-00041). Supported by the Spanish Ministry of Economy and Competitiveness (MINECO, MAT2014-51778-C2-1-R project, co-financed with FEDER). ICMAB acknowledges financial support from the Spanish Ministry of Economy and Competitiveness, through the "Severo Ochoa" Programme for Centres of Excellence in R&D (SEV- 2015-0496). The authors thank ICMAB Scientific and Technical Services. M.C acknowledges RyC contract 2013-12448.

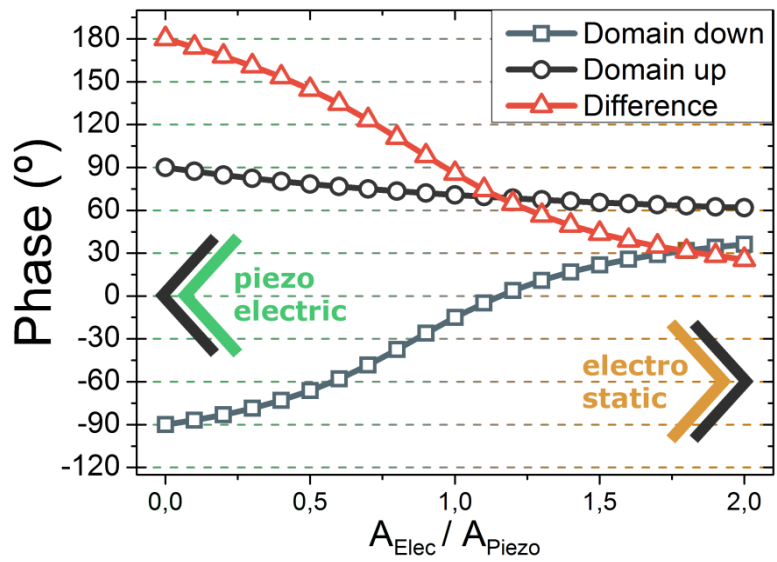

**Figure 1.** Theoretical representation of the total phase difference versus $A_{Elec}/A_{Piezo}$ for a sample with antiparallel domains. If no electrostatic forces are involved, the phase difference between antiparallel domains is 180º. As soon as electrostatic contribution appears, the total phase of the system decreases with the increased $A_{Elec}/A_{Piezo}$ ratio.

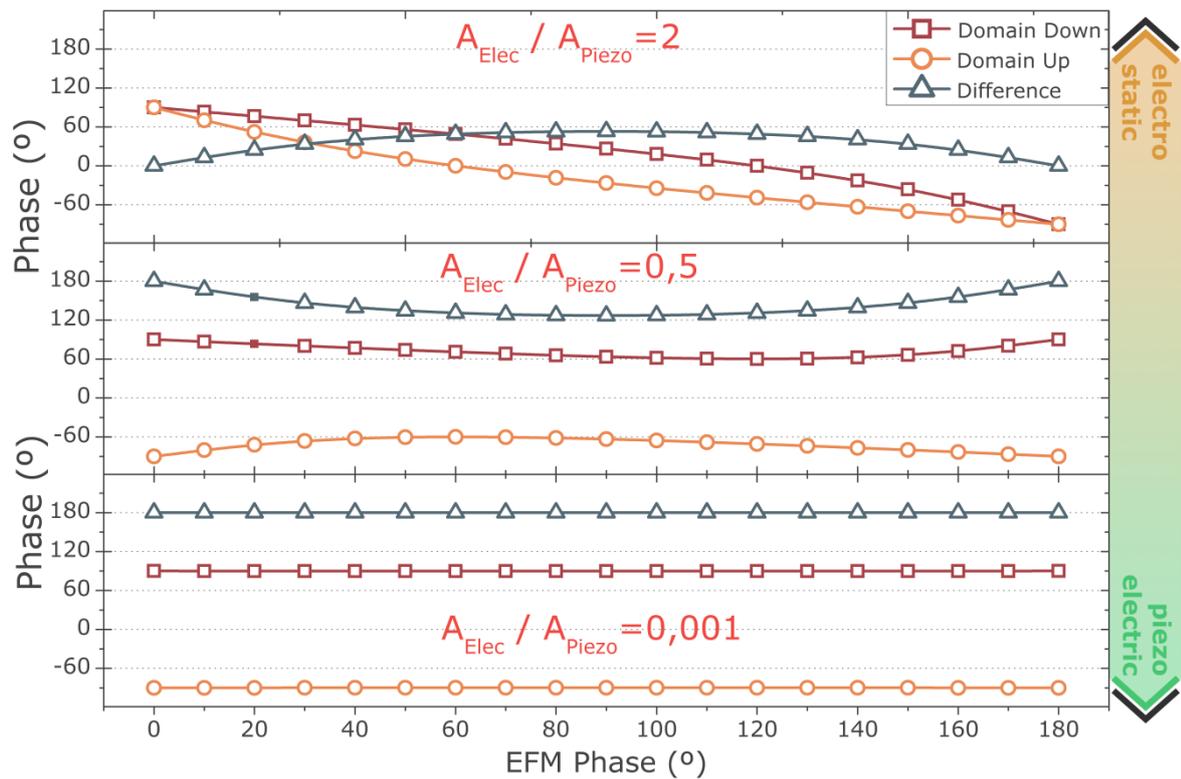

**Figure 2.** Theoretical description of the total phase difference versus the electrostatic signal phase, for different ratios of $A_{Elec}/A_{Piezo}$. For the case where the ratio is 0.001, the phase difference read is independent of the EFM phase signal. If this ratio increases, the phase for each of the domains decreases, up to a point that all the phase recorded is directly proportional to the EFM phase signal and independent of the piezoelectric signal.

| Tip name | k(N/m) | Sensitivity (nm/V) | Cantilever Length (µm) | Resonance Frequency(kHz) | Tip Material | Tip Length (µm) |
|---|---|---|---|---|---|---|
| NanoWorld EFM | 2,8 | 120,5 | 225 | 75 | PtIr | 15 |
| NanoWorld CDT-FMR | 6,2 | 257,1 | 225 | 105 | Diamond | 15 |
| AppNano DDSICONA | 0,2 | 630,5 | 450 | 12 | Diamond | 15 |
| RockyM 25PT400 | 8 | 384,6 | 400 | 10 | Solid Pt | 80 |
| RockyM 25PT300 | 18 | 561,6 | 300 | 20 | Solid Pt | 80 |
| RockyM 25PT200B-H | 250 | 248,2 | 200 | 100 | Solid Pt | 80 |

**Table 1:** Specifications comparison between different conductive tips available in the market

**Table 1.** Comparison between the different conductive tips used to carry on PFM measurement which are commercially available.

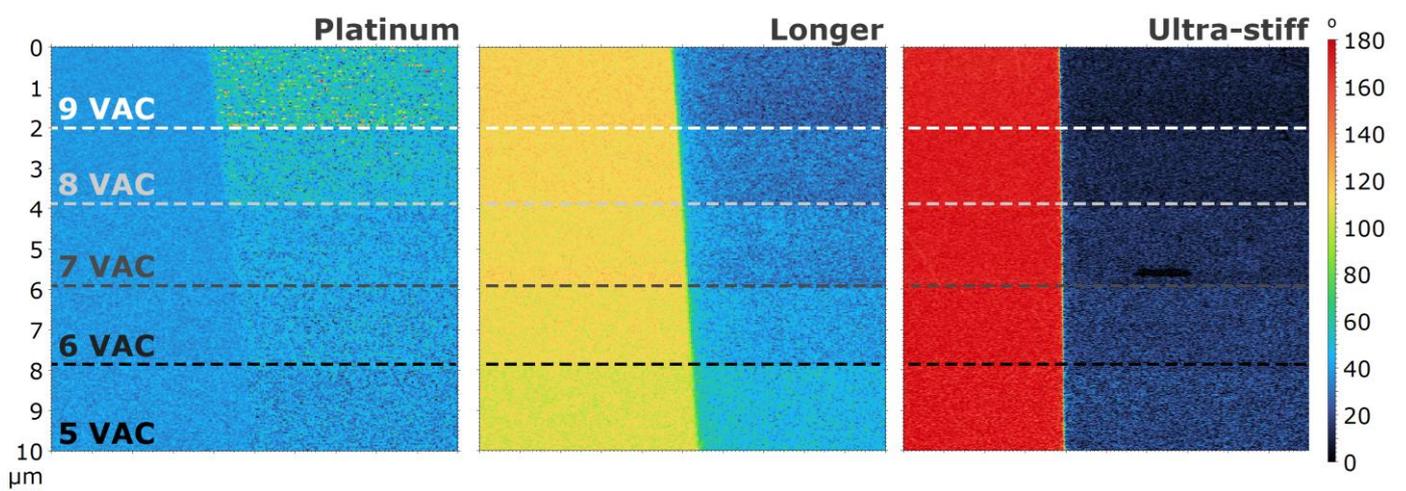

**Figure. 3**. From left to right, PFM Phase images of two antiparallel domains of a PPLN, obtained with a platinum coated tip (NW EFM), a longer solid platinum tip (RMN-25PT400) and ultra-stiff tip (RMN-25PT200H). 105 kHz AC voltage was applied to the tip, the amplitude was increased from 5VAC to 9VAC, the amplitude was increased at a pace of 1V each 2 microns

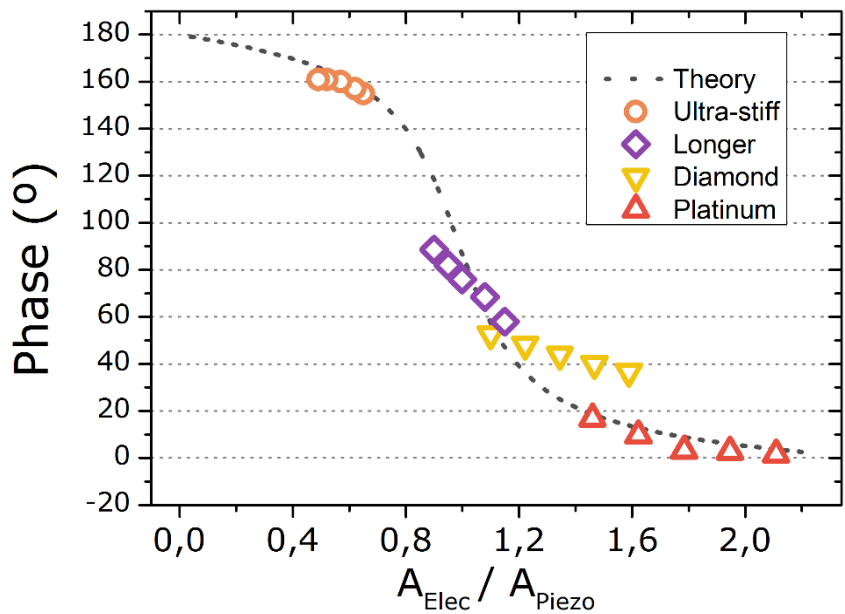

**Figure 4.** Phase difference between up and down domains versus $A_{Elec}/A_{Piezo}$ ratio. Dotted line corresponds to the theoretical prediction while experimental data is fitted into the curve, corresponding to the ultra-stiff, longer, diamond and platinum coated tips. It is seen that the less electrostatically influenced tip is the ultra-stiff, followed by longer tips.

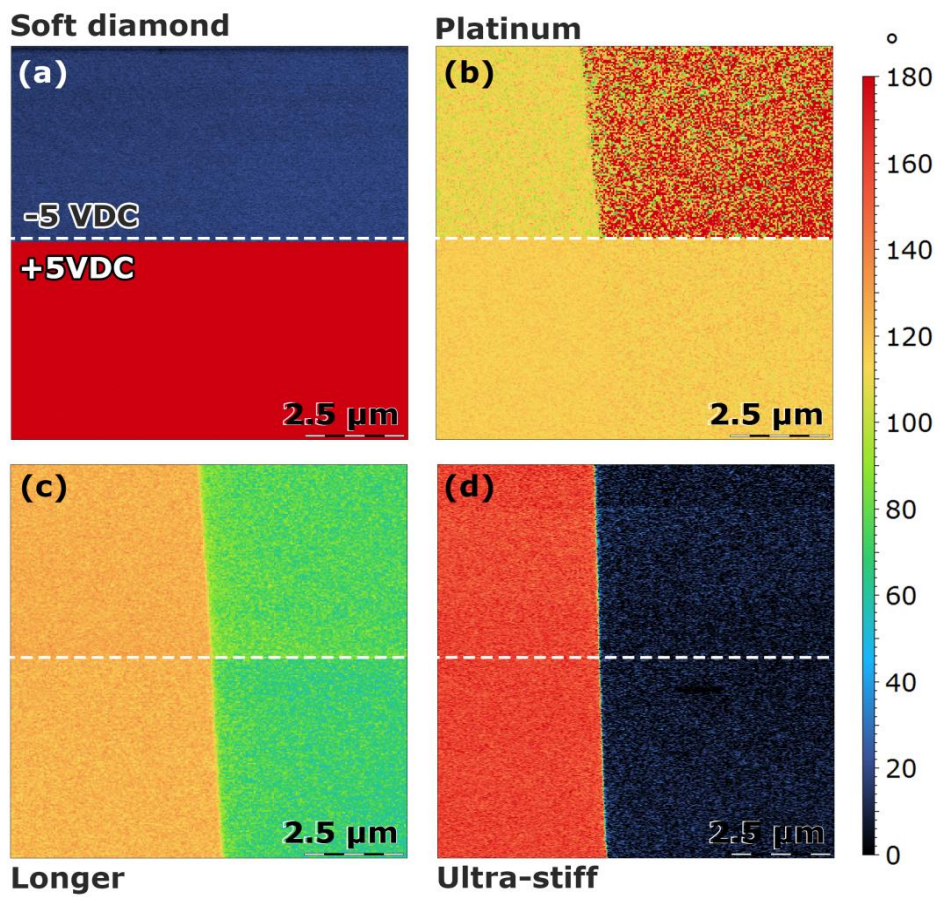

**Figure 5.** PFM phase image obtained with a soft diamond coated probe (a), a standard platinum coated tip (b), a longer tip length probe (c) and a ultra-stiff probe (d). Image acquired with 5Volts AC voltage, 105kHz frequency, +5VDC was applied to the sample from bottom to the middle, while -5VDC is applied from middle to bottom. It is found that both soft diamond and platinum tip are extremely sensitive to DC bias, while longer ultra-stiff tips are independent of such applied DC bias.